\documentclass[
superscriptaddress,
nofootinbib,
amsmath,amssymb,
aps,
pre,
longbibliography,
notitlepage,
showkeys
]{revtex4-1}
\usepackage[left=2.5cm,top=3cm,right=2.5cm,bottom=3cm,bindingoffset=0.5cm]{geometry}
\usepackage{graphicx}
\usepackage{hyperref}
\usepackage[bitstream-charter, greekfamily=default]{mathdesign}
\usepackage{lipsum}
\usepackage{comment}
\usepackage{braket}
\usepackage{rotating}
\usepackage{booktabs}    
\usepackage{physics}
\usepackage[dvipsnames]{xcolor}
\usepackage{inputenc}[utf-8]
\usepackage{tikz-network}
\usetikzlibrary{matrix}
\usepackage{natbib}[squares, numbers]

\newcommand{\avg}{\mathbb{E}}
\begin{document}
\title{Revealing production networks from firm growth dynamics} 
\author{Luca Mungo}
\affiliation{Mathematical Institute and Institute for New Economic Thinking at the Oxford Martin School, University of Oxford, Oxford, United Kingdom}
\email{luca.mungo@maths.ox.ac.uk}
\author{Jos\'{e} Moran}
\affiliation{Mathematical Institute and Institute for New Economic Thinking at the Oxford Martin School, University of Oxford, Oxford, United Kingdom}
\affiliation{Complexity Science Hub Vienna, Josefst\"adter Stra{\ss}e 39, A-1080, Austria}
% \email{jose.moran@maths.ox.ac.uk}

\begin{abstract}
    We study the correlation structure of firm growth rates. We show that most firms are correlated because of their exposure to a common factor but that firms linked through the supply chain exhibit a stronger correlation on average than firms that are not. Removing this common factor significantly reduces the average correlation between two firms with no relationship in the supply chain while maintaining a significant correlation between two firms that are linked. We then investigate if this observation can be used to reconstruct the topology of a supply chain network using Gaussian Markov Models.

    % Modern firms are tangled in complex, global production networks; Understanding these networks is crucial to several objectives, including predicting the propagation of shocks and computing goods' carbon footprints. Nevertheless, these networks are largely unmapped, as fine-grained data on firms' commercial counterparts is scarcely available. In this paper, we study how production networks impact firms' growth. We use a large dataset of firms' financials and supply-chain information to show that, on average, the closer two firms are on the supply chain, the more correlated their growths will be. We then demonstrate how this observation can be used to reconstruct the topology of a supply chain network using Gaussian Markov Models. 
\end{abstract}

\maketitle

% Main body starts here
\section*{Introduction}
\label{sec:intro}

Fifty years ago, Wassily Leontief was awarded the Nobel prize in Economics for his \textit{development of the input-output method and its application to important economic problems}.\footnote{
    \href{https://www.nobelprize.org/prizes/economic-sciences/1973/summary/}{See e.g. https://www.nobelprize.org/prizes/economic-sciences/1973/summary/}. Interestingly, Leontief took inspiration from the \textit{tableau économique}~\cite{quesnay} of the physician-turned-economist Quesnay, a member of the physiocratic school of economic thought, which saw the economy much like a human body. To Quesnay, mapping the relationships within the economy was equivalent to studying human anatomy.
}. His input-output framework~\cite{Leontief1936} views industries as nodes in a network of physical and monetary flows. Conservation laws for these flows lead, at economic equilibrium, to linear systems of equations linking the production of different industries, whose solutions show how differences in the output of an industry impact the output of any other economic sector. 

These equations were used to determine, for example, how much one should invest in each sector of an economy to increase the production of a given sector.\footnote{This was indeed not an easy problem: to increase the production of steel, it is necessary to increase the production of coal, but coal extraction requires having steel. Input-output analysis provided tools to solve this conundrum.}. It was in particular an important tool for central planners in the decades following the Second World War~\cite{bollard2019}. 

Later on, input-output analysis was used to understand the origins of macroeconomic fluctuations, with the seminal paper of Long and Plosser~\cite{long1983real}, where the input-output network amplifies small shocks that can lead to system-wide crises. However, most of these analyses are conducted at a very coarse-grained level, in the sense that they attempt to model the different \textit{sectors} of the economy rather than modeling more granular constituents: there are 405 industries in U.S. Bureau of Economic Analysis' most disaggregated input-output tables, while there are approximately 200 million firms worldwide. This is an unsettling remark, as recent literature ~\cite{acemoglu2012network, carvalho2020, diem2022quantifying} shows that fine-grained production networks play an important role in the propagation of shocks and that aggregating firms into sectors can lead to a misestimation of risk and distress propagation. Detailed firm-level data will also be crucial to the coming of age of agent-based modeling, a promising approach to studying \textit{out-of-equilibrium} macro-economic phenomena~\cite{Dessertaine2022}, that recently matched the forecasting accuracy of more traditional methods~\cite{Poledna2023, canvas, pichler_covid}.

Firm-level production data is thus very useful, but is also scarce~\cite{bacilieri2022}: the few datasets that are available only cover certain countries or certain categories of companies, leaving most of the global production network inaccessible. To tackle this problem, recent efforts have attempted to \textit{reconstruct} the production network, inferring the topology of the network using only partial, aggregate or related data. For instance, ~\cite{reisch2022inferring} uses mobile phone data to reconstruct the supply chain network of an undisclosed European country, while ~\cite{Brintrup2018} and ~\cite{Brintrup2021} pioneered machine learning for link prediction in supply chains, leveraging topological features computed by hand or distilled automatically through Graph Neural Networks. A similar approach was used in~\cite{mungo2022} to predict links between firms using their financial, industrial, and geographical features. Additional efforts have been carried out to adapt maximum-entropy models ~\cite{Hooijmaaijers2019,Mattsson2021,Ialongo2022, SquartiniEtAl2018, EnhancedGravity,SquartiniGarlaschelli2011, SquartiniEtAl2015}, already popular for models of international trade ~\cite{Gravity1, Gravity2, Gravity3, Gravity4, EnhancedGravity}, to the reconstruction of firm-level networks. The motivation of this research effort is that economic models conceived to represent the economy at the firm level require a good knowledge of the production network and should lead to a better understanding of economic dynamics and forecasts. But the converse should also be true: supply chains are vital in a firm's production, and they should leave a trace on the dynamics of a firm, something that has been observed when considering natural disasters~\cite{carvalho2020} or the dynamics of companies' market capitalisation~\cite{Abergel2022}. Is it possible to work backward from this, and infer the network topology from firm dynamics?

The study of firm dynamics, through the statistical analysis of their growth rates, has a long history dating back to the work of Gibrat~\cite{gibrat_legacy}. Gibrat's model is a multiplicative growth model initially proposed to explain the distribution of firm sizes (proxied, e.g., by sales or their number of employees). The model assumes that a firm grows by a random percentage of its current size from one period to the next. This random variable is thought of as being independent across firms and was initially also modeled as having the same distribution for all companies. Although this last hypothesis has been weakened in past work, showing for example that the volatility of firm growth decreases with their size in a non-trivial way~\cite{amaral1997scaling}, and even that it is necessary to think of the volatility of growth as being firm-dependent~\cite{growthrates}, the hypothesis of independence has not been explicitly questioned thus far. We propose to go beyond this, making the dependencies between firm growth explicit by studying the correlations between them and leveraging this information to reconstruct the firm network.

This paper is organized as follows. Section \ref{sec:data} gives an overview of the data we use for our paper, which we use in conjunction with the methods we outline in Section \ref{sec:growth_time_series}. Section \ref{sec:network_correlation} presents clear empirical evidence of the link between the supply chain and firm growth. Section \ref{sec:reconstruction} makes use of these observations and attempts to reconstruct the production network from firm growth time series. We detail both the optimization algorithm used to carry out this reconstruction as well as the results we obtain. Finally, Section \ref{sec:conclusions} concludes.

\section{Data}
\label{sec:data}

The primary data sources used in this article are the FactSet Fundamentals and FactSet Supply Chain Relationships datasets. Together, they provide a coherent environment from which companies' financial information (such as their quarterly sales or market capitalization), legal information (e.g., their industrial classification or headquarters location) and supply chain connections can be retrieved. Although it is very large, it should be noted that this dataset has a strong bias in covering mainly US firms.

The first dataset contained in this environment, FactSet Fundamentals, contains firms' financial, balance sheet, and legal information. The dataset spans a time range going from the early 1980s to the present day and covers developed and emerging markets worldwide for a total of around $100, 000$ active and inactive companies. From 1995 onwards, data on firms' sales, capitalization and investments is available for each quarter.

The second dataset, FactSet Supply Chain Relationships, is assembled by FactSet using multiple sources. The most prominent of these are filings required by the US Federal Accounting Standards, whereby each firm must report its most important suppliers and clients, and import-export declarations from bills of lading. These sources are complemented with insight mined by FactSet from news, press releases, company websites, and other sources of business intelligence, which permit the inference of a link between two companies.
Each record of a link between two companies can be represented by a temporal network, using directed links connecting a supplier to its customers. The temporal dimension of this data is also provided by FactSet: each link is assigned specific timestamps indicating the first time the connection was reliably attested and when the connection is known to have ended, when this is the case.\footnote{Note that this procedure implies that persistent links appear multiple times, as they are reported over many years.}

To simplify our analysis, we have discarded the temporal dimension by aggregating all the links into a single network that only considers whether a link between two companies was ever present in the time period we consider. Another simplification we perform is to aggregate firms that may be part of large conglomerates at the ultimate parent level using ownership structure data. Thus, the total sales, market capitalization and any other balance sheet data of these aggregated entities are the sum of these quantities for each of the constituting entities. At the network level, this procedure has the effect of deleting possible self-loops, as, for example, two branches of the same conglomerate that are present in separate countries can trivially be reported to have supply chain linkages between them. These aggregated entities constitute what we understand by ``firms'' or ``companies'' in the remainder of this paper.

Finally, we have only retained firms in the global supply chain's \textit{weakly largest connected component}\footnote{
A weakly connected component is a set of nodes such that for any two nodes $A$ and $B$, there exists a directed path starting at $A$ and arriving at $B$ or from $B$ to $A$, but not necessarily the other way around. When both a path $A\to \ldots\to B$ and $B\to \ldots\to A$ exist for any two nodes $A$ and $B$ in the component, a much more restrictive condition, then it is said to be strongly connected.
}, whose financial information was available for at least eight years.\footnote{The reason for this is to remove time series that are too short for our analysis, as the reader will appreciate later.} 
Our final sample is composed 
%\JM{Leaving both global and US numbers here / UPDATE: needs UPDATING}
of $16, 401$ firms connected by $178,911$ links. Appendix \ref{app:dataset_construction} details how to transform FactSet's original tables into our working dataset.

\begin{table}[h]
    \centering
    \begin{tabular}{r | c}
        \toprule
        \midrule
        Number of firms & $16,401$ \\
        Number of links & $178,911$ \\
        Density & $6.7\times 10^{-4}$\\
        Median degree & $7$ \\
        Max. degree  & $1664$\\
        \bottomrule
    \end{tabular}
    \caption{Network summary statistics}
    \label{tab:networkstats}
\end{table}

\section{Growth time series}
\label{sec:growth_time_series}

We label firms with an index $i=1,\ldots, N$, calling $s_i(t)$ and $m_i(t)$ the sales and market capitalization (the stock price multiplied by the number of shares outstanding) of firm $i$ at time $t$ (counted in quarters). With this, we define the annual growth rate of the sales of the firm as
\begin{equation}
    g_i(t) := \log\left(\frac{s_i(t+4)}{s_i(t)}\right).
\end{equation}
This quantity describes sales variations over the scale of a year, sampled with a quarterly frequency. %At this yearly scale, it is well known that log-returns are approximately Gaussian ~\cite{plerou1999scaling}. 
We follow Ref.~\cite{growthrates} in describing sales growth rates with a random variable with a Gaussian central region, although with fatter tails than a normal distribution, along with firm-dependent mean and variance (volatility). This therefore leads us to define the rescaled growth rates,
%
% From a preliminary study ~\cite{growthrates}, we know that these quantities have a firm-dependent mean and variance (volatility), and also that they have fatter tails than a Gaussian distribution. In accordance, we use the \textit{leave-one-out} rescaling described in ~\cite{Bouchaud2003}, leading us to define the rescaled growth-rates and returns,
%
\begin{equation}\label{eq:resc}
\begin{split}
	g'_i(t) := \frac{g_i(t) - \mathbb{E}_{t'}\left[g_i(t')\right]}{\sqrt{\mathbb{V}_{t'\neq t}\left[g_i(t')\right]}}\\
\end{split}
\end{equation}
where the average is computed over all times $t'$, but the variance is computed from the time series where the observation corresponding to $t'=t$ has been removed. This corresponds to the \textit{leave-one-out} rescaling defined in \cite{Bouchaud2003}, where the denominator on the right-hand side of Eqs.\eqref{eq:resc} allows one to rescale with respect to the volatility when considering a variable with a fat-tailed distribution.\footnote{Indeed when the distribution is fat-tailed then the naive estimator for the variance, related to $\sum_t g_i(t)^2$, may be dominated by a single observation (the largest one in the sample) and therefore introduce an artificial cut-off when dividing by the variance because in this case $\sum_i g_i(t)^2 \approx \max_t g_i(t)^2$. When rescaling the largest value in the sample, it is clear that it may be clipped because of this.} We drop the apostrophe below for clarity, as we will not use the ``bare'' growth rates in the remainder of this article.

Our goal in the rest of this article is to infer the supply chain structure from the correlation structure of the growth rates.  Nonetheless, it is likely that the growth rates of two companies are correlated because of reasons other than their connection through the supply chain. This can be the case, for instance, if two firms are in a given country that endures an exogenous economic shock, as in the case of the Covid-19 pandemic. Our strategy therefore will be to attempt to remove these common factors, assuming that what remains in the correlations must be the more subtle effects due to the supply chain. To illustrate the technique used for this, we shall resort to a very simple model that is described below.

\subsection{Removing common shocks}
\label{ssec:simple_example}

Let us propose first a very simple example, where one has $N$ time series $x_i(t)$, with $1\leq i \leq N$ and $1\leq t \leq T$. Each time series $x_i\left(t\right)$ is composed of an idiosyncratic term, driving time series $i$ only and given by i.i.d. Gaussian terms, and a common term that affects all the time series and that is also random. The model reads
\begin{equation}\label{eq:common_mode_model}
    x_i(t) = \xi_i(t) + \sigma v(t),
\end{equation}
where $\xi_i(t)$ is a Gaussian random variable with $\mathbb{E}[\xi_i(t)] = 0$ and $\mathbb{E}[\xi_i(t)\xi_j(t')]= \delta_{ij}\delta_{tt'} $, with $\delta_{ij}$ the Kronecker delta (i.e., $\delta_{ij}=1$ if $i=j$ and $0$ otherwise). Similarly, $v(t)$ is a Gaussian random variable satisfying $\mathbb{E}[v(t)v(t')]=\delta_{tt'}$ and $\mathbb{E}[v(t)\xi_i(t')] = 0$.

In this case, where we know precisely the nature of the common shock, we can estimate $v(t)$ when $N$ is large by writing:
\begin{equation}\label{eq:estimation}
    \frac{1}{N} \sum_{i=1}^N x_i(t) = \frac{1}{N} \sum_{i=1}^N \xi_i(t) + \sigma v(t) \underset{N\gg 1}{\approx} \sigma v(t).
\end{equation}
The correlation matrix for the model's time series reads
\begin{equation}\label{eq:corr_fct_common_mode}
\begin{split}
    C_{ij} &:= \mathbb{E}[x_i(t)x_j(t)]
    =  \delta_{ij} + \sigma^2,
\end{split}
\end{equation}
which we can rewrite as $\mathbf{C} = \mathbf{I} + N\sigma^2 \mathbf{u} \mathbf{u}^\intercal$, with $\mathbf{u}=\frac{1}{\sqrt{N}}\mathbf{1}$, and where $\mathbf{u}^\intercal$ indicates vector transposition.\footnote{This vector $\mathbf{u}$ is chosen to be normalised.} 
Because $\mathbf{C}$ is the sum of the identity matrix and a rank-one matrix, it is easy to see that it has an eigenvalue $1+\sigma^2$, corresponding to the eigenvector $\mathbf{u}$ as $\mathbf{C}\mathbf{u} = (1+N\sigma^2)\mathbf{u}$, with all the other $N-1$ remaining eigenvalues equal to $1$, with eigenvectors corresponding to the canonical basis of the vector space that is orthogonal to $\mathbf{u}$.
We can in fact go further in this geometric interpretation and bring meaning to the vector $\mathbf{u}$ by focusing on the \textit{projection} of the time series onto it. What we mean by this is that for every time step in the multi-dimensional time series, we may consider the vector $\mathbf{x}(t)=\left( x_1(t),\ldots, x_2(t) \right)$, and consider the projected time series $\hat{v}(t)= \mathbf{u}\cdot \mathbf{x}(t)$. 

In this case, we notice that for large $N$ we should have $\hat{v}(t)= \frac{1}{N}\sum_{i=1}^N x_i(t)\approx \sigma v(t)$. We can actually generalize this: if we replace Eq.\eqref{eq:common_mode_model} by
\begin{equation}\label{eq:common_mode_gen}
    x_i(t) = \xi_i(t) + \sigma u_i v(t),
\end{equation}
that is a model where each time series has a different exposure (or loading, in factor-models' jargon) to the common mode $v(t)$, then the correlation matrix is the same and we still have an eigenvector $\mathbf{u}=(u_1,\ldots,u_N)$.\footnote{This vector can be assumed to be normalised, if not we can always replace $\sigma$ by $\sqrt{\mathbf{u}^2}\sigma$ in the model.} Doing the projection $\mathbf{x}(t)\cdot \mathbf{u}(t)$ still leads to $\hat{v}(t)\approx v(t)$.

% We can thus diagonalise the correlation matrix, and write $\mathbf{C} = \mathbf{PDP}^\intercal$, with 

% \begin{equation}\label{eq:diagonalising_matrix}
%     \left\{\begin{matrix}
%         P_{i1} &=& \frac{u_i}{\sqrt{\mathbf{u}^2}} &=& \frac{1}{\sqrt{N}}\\
%         P_{ij} &=& \sqrt{\frac{N}{N-1}}\left(\delta_{ki}-\frac{1}{N}\right), & & 1< j \leq N
%     \end{matrix} \right.
% \end{equation}
% and $\mathbf{D}=\text{diag}(1,\ldots,1,N\sigma^2)$. This matrix $\mathbb{P}$ is obtained by imposing that its first column is given by $\mathbf{u}/\sqrt{\mathbf{u}^2}$ and getting the other columns by using the Gramm-Schmidt orthonormalisation procedure to obtain an orthonormal basis. 

In fact, we can also consider the \textit{orthogonal projector} to $\mathbf{u}$, given by $\mathbf{P}=\mathbf{I} - \mathbf{u}\mathbf{u}^\intercal$, or equivalently $P_{ij} = \delta_{ij}-u_{ij}$. We can now apply this projector to our time series, as $\mathbf{y}(t) = \mathbf{P}\mathbf{x}(t)$, or equivalently by defining $\mathbf{Y}=\mathbf{P}\mathbf{X}$. It is straightforward to check that $y_i(t) = x_i(t) - \hat{v}(t) \approx \xi_i(t)$.

To address our general problem of removing common fluctuations from time series, we can adopt the following procedure to remove the common mode and be left only with the idiosyncratic fluctuations. Assuming that the common mode $v(t)$ is the primary driver of time series variations ($\sigma \gg 1$), we can:

\begin{enumerate}
    \item Take the time series and compute the empirical correlation matrix,
    \item Diagonalise the correlation matrix and rank the eigenvalues and eigenvectors according to the magnitude of the eigenvalue,
    \item Project the time series onto the eigenvector corresponding to the largest eigenvalue to get the dynamics of the common mode,
    \item Remove the dynamics of the common mode from the time series by using the orthogonal projector to the corresponding eigenvector.
\end{enumerate}

\begin{figure}
        \includegraphics[width=\textwidth]{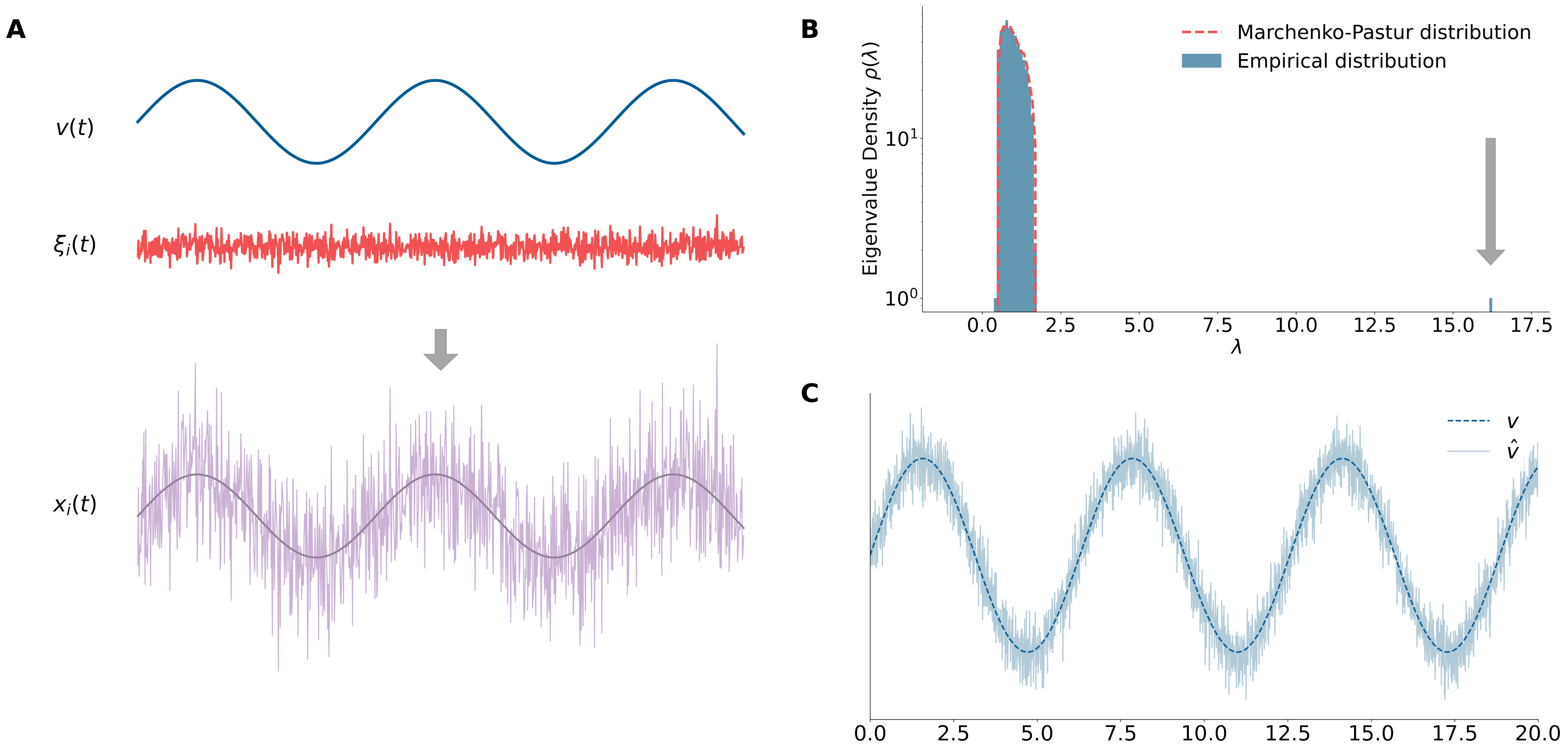}
        \caption{(A) The time series $x_i\left(t\right)$ are created by adding a sine wave and an idiosyncratic random noise. (B) The spectrum of the empirical correlation matrix $\widehat{C}_{ij} = \frac{1}{T}\sum_{t=1}^Tx_i\left(t\right)x_j\left(t\right)$, along with the random benchmark given by the Mar\v{c}enko-Pastur distribution. Note the presence of an eigenvector at $\lambda\approx 16$, beyond the random benchmark (C). The eigenmode $\hat{v}(t)$, obtained by projecting the time series onto the vector $\widehat{\mathbf{u}}$ corresponding to the largest eigenvalue, tracks the collective oscillations of the system.}
        \label{fig:time_series_example}
\end{figure}

Naturally, we can repeat this procedure and remove also the mode corresponding to the second largest eigenvalue and so on, so that it is easily generalizable to other, more complex situations than the one of Eq.\eqref{eq:common_mode_model} (see Fig. \ref{fig:time_series_example} for an example where the common mode $v(t)$ is a sinusoidal wave).

The issue, however, is that this relies on the assumption that the empirical correlation matrix is a reliable estimator of the ``true'' underlying correlation matrix from which the data is generated.\footnote{At least in this model. In reality, when analyzing time series with this point of view we are making the more stringent assumption that the correlation structure of data is time-invariant. Although there has been some work to relax this assumption in e.g. financial data~\cite{challet}, these approaches are difficult, if not impossible, to adapt to the time series we analyze because of their relatively small length and sampling frequency.} Naturally, this is not true, and one expects some estimation error when the length of the time series $T$ is finite. In our toy model above, it is in fact possible to separate the contribution of the idiosyncratic noise, as $\widehat{\mathbf{C}_0}:=\frac{1}{T}\left(\boldsymbol{\xi}\boldsymbol{\xi}^\intercal\right)_{ij}=\frac{1}{T}\sum_{t=1}^{T}\xi_i(t)\xi_j(t)$. Because the elements of $\boldsymbol{\xi}$ are i.i.d. Gaussian random variables, this empirical correlation matrix is known as a Wishart matrix~\cite{wishart}, and the statistical properties of its spectrum are known to be determined by the Mar\v{c}enko-Pastur distribution~\cite{Marcenko1967}. For a more in-depth understanding of this and other links with random matrix theory, we invite the reader to consult~\cite{Potters2020}, but we will explain the main results we need below.

 Because $\widehat{\mathbf{C}_0}\underset{T\to\infty}{\longrightarrow}\mathbf{I} $, we expect naturally that for large time series the spectrum of $\widehat{\mathbf{C}_0}$ should be concentrated around $1$. In practice, however, because of measurement error, we don't expect \textit{all} of its eigenvalues to be equal to $1$. Thus, we intuitively expect the full spectrum of $\widehat{\mathbf{C}}$ to be constituted of $N-1$ eigenvalues close to $1$, which constitute the contribution coming from $\mathbf{C}_0$, and a single-peaked eigenvalue close to $\sigma^2$, which is the contribution coming from the dynamics of $v(t)$ that couples all of the $N$ time series. For the full empirical correlation matrix $\widehat{\mathbf{C}}$, we also expect that the eigenvector corresponding to its largest eigenvalue will satisfy, $\widehat{\mathbf{u}}\approx \mathbf{u}$. However, the result of Mar\v{c}enko-Pastur is that in the limit where both $N, T \to \infty$, but with the ratio $q=\frac{N}{T}$ fixed, the spectrum of $\mathbf{C}_0$ is concentrated in the interval $(1-\sqrt{q}, 1+\sqrt{q})$, called the ``bulk'', and may also have a delta-peak at $0$ if $q<1$. For finite $N, T$ we also expect some eigenvalues to be slightly out of this interval. This sheds light on why in practice finding the common mode may be difficult: if, say, $\sigma$ is of the order of $q$, then the eigenvalue ``spike'' at $1+\sigma^2$ will in fact be inside the Mar\v{c}enko-Pastur interval. This is linked to the so-called Baik-Ben Arous-Péché (BBP) transition~\cite{Baik2005}, and in this case, it is not possible to reconstruct the common mode. 

We can indeed imagine that we run the model and execute the procedure described above first for a value of $\sigma\gg q$, and then reduce $\sigma$ progressively until we reach $\sigma \approx q$. When diagonalizing the empirical correlation matrix $\widehat{\mathbf{C}}$ and considering the eigenvector corresponding to its largest eigenvalue, $\widehat{\mathbf{u}}$, this eigenvector will match the ``true'' eigenvector $\mathbf{u}$ when $\sigma \gg q$, so that for example $\widehat{\mathbf{u}}\cdot \mathbf{u} \approx 1$. However, as $\sigma\to q$ this overlap will decrease, and the intuition then is that when the outlier eigenvalue reaches the Mar\v{c}enko-Pastur bulk, then its associated eigenvector $\widehat{\mathbf{u}}$ cannot now reliably be thought of as an estimator of $\mathbf{u}$, and will instead point in any random direction. In this case $\mathbf{u}\cdot\widehat{\mathbf{u}}$ will be of order $1/\sqrt{N}$ (see~\cite[Section 14.2.2]{Potters2020}, and also~\cite{bun-allez} for intuition for this phenomenon using Dyson Brownian motion). In this case, the usage of the projectors, or steps 3 and 4 of our procedure, will not lead to the identification of common modes.

The conclusion from this is that we are indeed capable of identifying common factors in time series using this approach, but we must first make sure that these modes correspond to eigenvalues of the correlation matrix that are not compatible with a random benchmark. 
Indeed, the example above corresponds to time series of equal length, where each entry of the time series is drawn at random from a Gaussian distribution. In this case, the random benchmark for the spectrum is determined by the Mar\v{c}enko-Pastur distribution, as said above. The case of our time series is, however, different since sales data is not available for every company at any time. Growth time series can have different starting points and lengths, and the period over which one can compute their correlation is different for any pair of firms. Our data therefore has a lot of missing values, and two firms present in non-overlapping times for example will be set two have a correlation of $0$. Another issue is that the growth-rate distribution is not Gaussian, and has slightly heavier tails. Understanding the correlation spectrum of heavy-tailed processes is feasible (see for example~\cite{Biroli2007}), but very difficult to do for any distribution.

We can nonetheless establish a random benchmark for the correlation spectrum computationally and use it to identify eigenvalues indicating correlated modes. We achieve this by creating a surrogate of the growth-rate time series where the missing data structure is preserved and where the individual growth rates are drawn at random from their empirical distribution. This is similar to the procedure used in~\cite{Vodenska2016}, where the authors randomly shuffle a time series to benchmark the eigenvalues of correlation matrices that can be distinguished from noise.

\begin{figure}
\includegraphics[width=\textwidth]{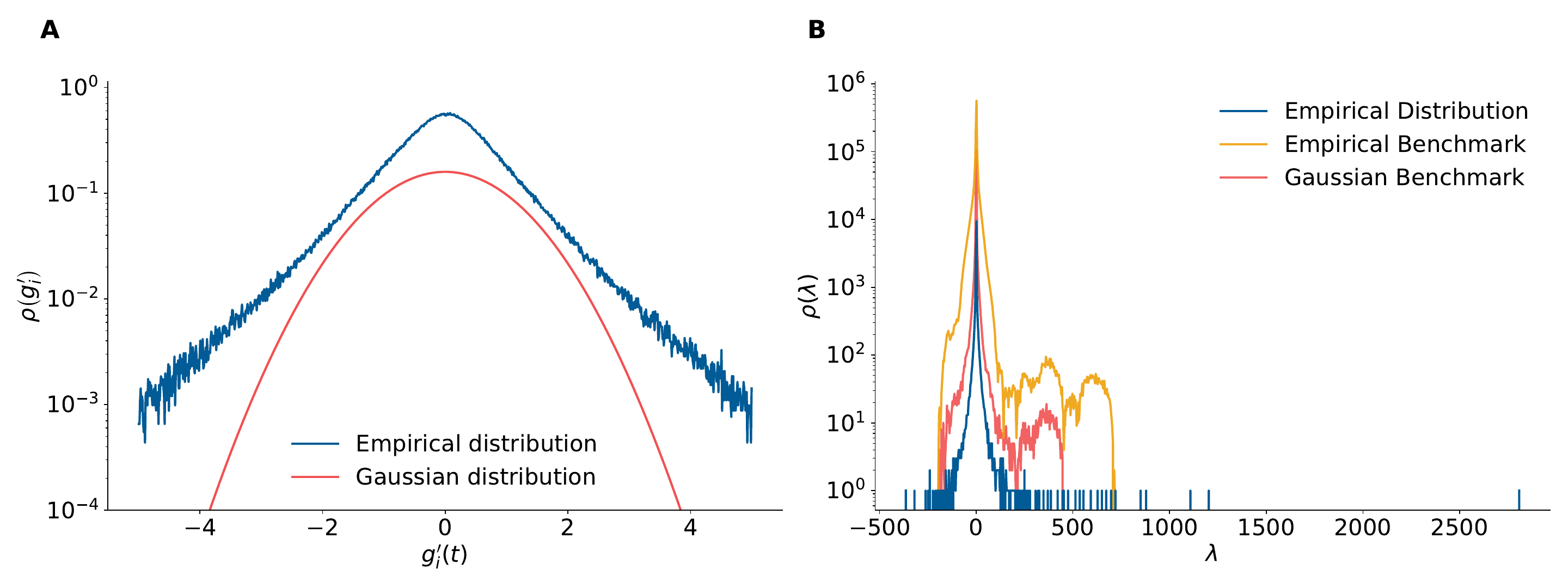}
\caption{(A) The distribution $\rho\left(g\right)$ of the growth rates for every firm $i$ and time $t$. A normal distribution is provided as a reference. (B) Growth time series correlation spectrum. The two random benchmarks are obtained by sampling random time series from the empirical distribution $\rho\left(g\right)$ (\textit{Empirical benchmark}) and the normal distribution (\textit{Gaussian benchmark}). The starting points and duration of the random time series match those of the real ones. The spectrum shown is the average of $10$ sets of random time series.}
\label{fig:distribution_and_spectrum}
\end{figure}

Figure \ref{fig:distribution_and_spectrum} shows that the real correlation spectrum has several eigenvalues that are beyond the bulk corresponding to the random benchmark, both on the left and on the right side of the bulk. Note that the presence of negative eigenvalues is a consequence of missing data, and is something that one does not obtain for standard Wishart matrices. The largest eigenvalue corresponds to the \textit{market mode}, a collective trend shared by all the firms in the supply chain. This collective mode concerns all firms, as shown by the fact that the entries of the corresponding eigenvector have (roughly) all the same sign and magnitude\footnote{This is similar to the toy model presented in Section~\ref{ssec:simple_example}.}. Thus, this mode corresponds to a common factor in the economy, and all the firms move coherently with it. Interpreting the modes corresponding to eigenvalues outside the bulk is more challenging: contrary to what is observed in the correlation structure of financial returns, we have not been able to identify them with specific industrial sectors or geographies. Because we are unable to give these eigenvectors a clear interpretation, and since they could potentially carry information about the production network, we have decided to remove only the first eigenmode from the time series. In the rest of our paper, we will refer to the growth time series cleaned of the system's first eigenmode as ``cleaned'' time series $\tilde{g}_i\left(t\right)$, and to their correlation as the ``cleaned'' correlation.\footnote{We attract the reader's attention to the fact that we mean ``cleaning'' in a sense that is the opposite of what is done for returns' correlation matrices in finance: there, usually one discards the modes corresponding to the \textit{smaller} eigenvalues (see e.g. \cite{bun2017}). We, however, discard the largest mode because we want to remove reasons for firm co-movement that are distinct from supply chain-induced co-movement.} 

{}

\section{Network correlation and random benchmarks}
\label{sec:network_correlation}
We have introduced the main object of our analysis, firms' growth time series $\mathbf{g}_i(t)$. We will now show that the supply chain induces specific correlations between firms, a necessary step to later justify our usage of correlations in supply-chain reconstruction. We define the following correlation matrices\footnote{Note that here we use the notation $\avg_t[\cdot]= \frac{1}{T}\sum_{t=1}^T \cdot(t)$ to indicate the empirical average across the time variable. The notation $\mathbf{E}$ used in the previous section corresponds instead to the ``true'' average value of our stochastic model, computed over the distribution of the noise $\xi_i$ and $v$. Similarly, $\mathbf{E}_{ij}$ indicates an empirical average taken by summing over the variables $i$ and $j$.},
\begin{equation}\label{eq:corr_functs}
\begin{split}
	C_{ij}(\tau) &= \avg_t\left[g_i(t) g_j(t+\tau)\right], \\
	\tilde{C}_{ij}(\tau) &= \avg_t\left[\widetilde{g}_i(t) \tilde{g}_j(t+\tau)\right]. \\
\end{split}
\end{equation}
We can compute the average value of the elements of the matrix $\mathbf{C}$ and $\widetilde{\mathbf{C}}$ across the pairs of firms $\left(i, j\right)$ linked in the production network, defining averaged client/supplier correlation functions. Given any (binary) adjacency matrix $\mathbf{A}$ we define
\begin{equation}
    C_{\mathbf{A}}\left(\tau\right) = \mathbb{E}_{ij} \left[C_{ij}\left(\tau\right)\vert A_{ij}=1\right],
    \label{eq:corr1}
\end{equation}
and
\begin{equation}
    \widetilde{C}_{\mathbf{A}}\left(\tau\right) = \mathbb{E}_{ij} \left[\tilde{C}_{ij}\left(\tau\right)\vert A_{ij}=1\right],
    \label{eq:corr2}
\end{equation}
where the average runs over all pairs $1\leq i \leq j\leq N$. In other words, $C_{\mathbf{A}}$ and $\widetilde{C}_{\mathbf{A}}$ are the average correlation between two neighbors in a graph with an adjacency matrix $\mathbf{A}$. This average can be computed using the \textit{true} adjacency matrix of the production network, $\mathbf{S}$, or over the adjacency matrix of any other network.

\subsection{Random benchmarks} % (fold)
\label{sub:random_benchmarks}
% subsection random_benchmarks (end)

We first compute the correlations averaged over the adjacency matrix $\mathbf{S}$ of FactSet's production network, where $S_{ij}=1$ if $j$ either supplies or is a client of $i$,  and compare their value to those obtained with several random network models: the \textit{Erd\H{o}s-R\'enyi} model \cite{ErdosRenyi}, the \textit{Stochastic Block Model} \cite{KarrerNewman2011}, and the \textit{Configuration Model} \cite{Newman2003}. We describe all three models and their parameters in detail below.

We randomly sample $n=50$ networks of each model, with adjacency matrices $\mathbf{R}_1, \ldots, \mathbf{R}_n$ and compute the mean and standard deviation of the sets $\left\{C_{\mathbf{R}_1}, \ldots, C_{\mathbf{R}_n}\right\}$ and $\left\{\tilde{C}_{\mathbf{R}_1}, \ldots, \tilde{C}_{\mathbf{R}_n}\right\}$. All of the models are parametrized to match the empirical properties of the supply-chain network.

For the Erd\H{o}s-Rényi network, we fix its density $p$ to match that of the production network, namely
$$
p = \frac{1}{N\left(N-1\right)}\sum_{i=1}^N\sum_{j>i}^N S_{ij}.
$$

The Erd\H{o}s-Rényi network has no real structure, and in particular no clear community structure is apparent in it. We therefore also used stochastic block models, which we initialised with several different block schemes. Specifically, we divided firms into blocks $\left\{B_1, \ldots, B_m \right\}$ depending on their industrial sector (at their SIC code's third-digit level of aggregation), their country, or their network community as identified by the Louvain community-detection algorithm \cite{louvain}. The network densities within- and across- blocks are chosen to be equal to their empirical counterparts,
\begin{equation}
\label{eq:blockmodels}
\rho_{ij} = \frac{1}{\left|B_i\right|\left(\left|B_j\right|-\delta_{ij}\right)}\sum_{u \in B_i, v \in B_j} A_{uv}.
\end{equation}
Finally, we use the configuration model to produce networks with a degree distribution that matches exactly the empirical one.

\begin{figure}[tb]
    \centering
    \includegraphics[ width=\textwidth]{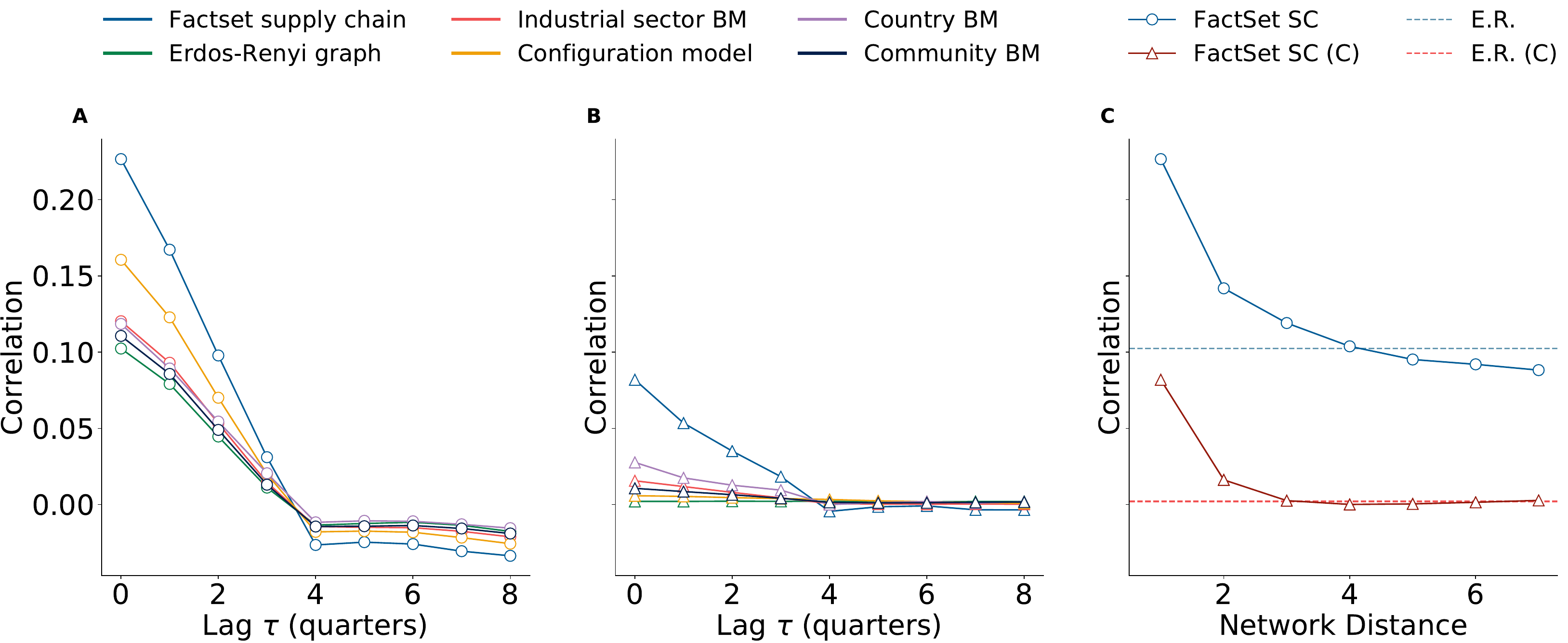} %trim={4.5cm, 0, 4.5cm, 0}, clip=true,
    \caption{(A): Average correlation on the production network $C\left(\tau\right)_S$ and several random network benchmarks. (B): Average "cleaned" correlation on the production network $\tilde{C}\left(\tau\right)_S$ and several random network benchmarks. (C): Correlations along the supply chain decay with distance. At distance $d=4$ ($d=3$ for the cleaned correlation), firms' average correlation is the same as the Erdos-Renyi benchmark. Results for the cleaned time series are flagged with a (C)}
    \label{fig:distance_corr}
\end{figure}

Figure \ref{fig:distance_corr} compares the average correlation measured on the true production network $\mathbf{S}$ and on the random network benchmarks. The value of $C_{\mathbf{S}}\left(0\right)$ is twice as high as the average correlation measured on the Erd\H{o}s-R\'enyi graph, and $\approx 50\%$ higher than the correlation measured for the configuration model. The result for $\tilde{C}_{\mathbf{S}}\left(0\right)$ are even more striking, with the residual correlation on the supply chain being still $\approx 0.1$ and most of the random benchmarks dropping close to zero. This highlights the usefulness of our cleaning procedure, as it significantly increases our signal-to-noise ratio.

\subsection{Relationship with network distance} % (fold)
\label{sub:relationship_with_network_distance}

% subsection relationship_with_network_distance (end)

A second way to show that the supply chain induces correlations in the dynamics of firm sales is to study how the correlation behaves with respect to network distance. Intuitively, we expect that two firms that are close to each other on the supply chain will be more correlated than two firms that are far apart.

To see this, we start again from the binary adjacency matrix $\mathbf{S}$ of the production network and define recursively
\begin{equation}
    S_{ij}^{(k)} = \sum_{l_1,\ldots,l_{k-1}}\mathbf{1}\left(S_{il_1}S_{l_1l_2}\ldots S_{l_{k-1}j}>0\right) \prod_{m=1}^{k-1} \left(1-S_{ij}^{(m)}\right),
\end{equation}

where $S_{ij}^{(1)}=S_{ij}$. The first factor in the right-hand side is equal to $1$ if and only if there exists a path $i\to l_1\to\ldots\to j$ of length $k$ linking $i$ to $j$. The second factor is $0$ if it exists a shorter path from $i$ to $j$ in the network. Thus defined, $S^{(k)}_{ij}$ is equal to one only if the shortest path between $i$ and $j$ is of length $k$.

\begin{figure}[tb]
    \centering
    \includegraphics[width=.5\textwidth]{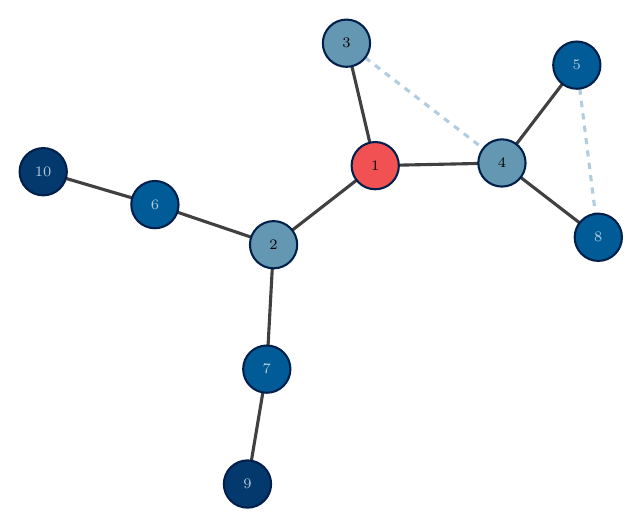}
    \caption{An illustration of network distance. Nodes 2, 3 and 4 are at a distance $k=1$ from node 10. Even though the path $1\to3\to4$ exists, we do not consider $4$ to be at distance $k=2$ from $1$}
\label{fig:node_distance}
\end{figure}
We can see how these correlations decay with distance, by computing the values

\begin{equation}
	D_{S}(k) = \avg_{ij}\left[C_{ij}(0)\vert S_{ij}^{(k)}=1\right],
\end{equation}
and
\begin{equation}
    \widetilde{D}_{S}(k) = \avg_{ij}\left[\tilde{C}_{ij}(0)\vert S_{ij}^{(k)}=1\right],
\end{equation}
namely the average of the non-lagged growth correlation between any two firms that are $k$-steps apart in the supply chain. We show this in Figure~\ref{fig:distance_corr}, C. The correlation between firms decays as their distance in the production networks increases, revealing again that the production network mediates growth correlations between firms.

\section{Supply Chain Reconstruction} % (fold)
\label{sec:reconstruction}

In the previous Sections, we have established that the supply chain induces correlations between firms, and we have also established that our cleaning procedure increases the signal-to-noise ratio of these correlations with respect to the real supply chain. We next propose a procedure to reconstruct the supply chain using the cleaned correlation matrix.

Inferring networks from observations or \textit{graph learning}~\cite{Dong_et_al_2019}, is a problem that encompasses several branches of natural and social sciences. Following \cite{Dong_et_al_2019}, we define the problem of graph learning as follows: given $T$ observations on $N$ entities, represented by a data matrix $\mathbf{X} \in \mathbb{R}^{N\times T}$,  and taking some prior knowledge as given, we seek to infer relationships between our $N$ entities and represent these relationships as a graph $\mathcal{G}$.

A possible approach to solve this problem is to assume that $\mathcal{G}$ encodes some statistical relationship between the entities. Specifically, \textit{probabilistic graphical models} assume that the structure of $\mathcal{G}$ determines the joint probability distribution of the observations on the data entities: the presence or absence of edges in the graphs encodes the conditional independence among the random variables represented by the vertices. In particular, \textit{Markov Random Fields} consider a graph $\mathcal{G}=\{ \mathcal{V}, \mathcal{E} \}$ and a set of random variables $\mathbf{x} = \left\{x_i : v_i \in \mathcal{V}\right\}$ satisfying the pairwise Markov property,
\begin{equation}
\label{eq:pairwise_markov_property}
	\left(v_i, v_j\right) \notin \mathcal{E} \Leftrightarrow p\left(x_i |x_j, \mathbf{x} \backslash\left\{x_i, x_j\right\} \right) = p\left(x_i, \mathbf{x} \backslash\left\{x_i, x_j\right\} \right),
\end{equation}
which simply states that two variables $x_i$ and $x_j$ are conditionally independent if there is no edge between the corresponding vertices $v_i$ and $v_j$. In Markov Random Fields, the joint probability distribution of the variables $x_1, \ldots, x_N$ may also be represented as
\begin{equation}
\label{eq: markov_random_fields}
p\left(\mathbf{x}\right) = \frac{1}{Z}\prod_{i=1}^K\phi_i\left(\mathbf{D}_i\right),
\end{equation}
where $\mathbf{D}_i, \ldots, \mathbf{D}_K$ are a set of graph's cliques (i.e., groups of nodes), $Z$ is a normalization factor known as the partition function, and $\phi_i$s are generic functions known as factors. It is straightforward to see that the exponential family of distributions with a parameter matrix $\mathbf{\Theta} \in \mathbb{R}$,
\begin{equation}
\label{eq:graphical_gaussian_models_1}
	p\left(\mathbf{x}|\mathbf{\Theta}\right) = \frac{1}{Z\left(\mathbf{\Theta}\right)}\exp\left(\sum_{v_i \in \mathcal{V}}\theta_{ii}x_i^2 + \sum_{(v_i, v_j) \in \mathcal{E}}\theta_{ij}x_ix_j\right),
\end{equation}
is compatible with this formalism; the multivariate Gaussian distribution with precision matrix $\mathbf{\Theta}$,
\begin{equation}
\label{eq:graphical_gaussian_models_2}
	p\left(\mathbf{x}|\mathbf{\Theta}\right) = \frac{\left|\mathbf{\Theta}\right|^{1/2}}{\left(2\pi\right)^{N/2}}\exp\left(-\frac{1}{2} \mathbf{x}^T\mathbf{\Theta}\mathbf{x}\right),
\end{equation}
belongs to this family. The subclass of Markov random fields that adopt Eq.\eqref{eq:graphical_gaussian_models_2} as the parametrization for the joint probability distribution $p$ are called Gaussian Markov Random Fields or Gaussian Graphical Models. 
%
\begin{comment}
	Suppose we have $N$ random variables, then this condition holds for the exponential family of distributions with a parameter matrix $\mathbf{\Theta} \in \mathbb{R}$,
	%
	\begin{equation}
	\label{eq:graphical_gaussian_models_1}
		p\left(x|\mathbf{\Theta}\right) = \frac{1}{Z\left(\mathbf{\Theta}\right)}\exp\left(\sum_{v_i \in \mathcal{V}}\theta_{ii}x_i^2 + \sum_{(v_i, v_j) \in \mathcal{E}}\theta_{ij}x_ix_j\right),
		\end{equation}
	%
	where $\theta_{ij}$ is the $ij$th entry of $\mathbf{\Theta}$, and $Z\left(\mathbf{\Theta}\right)$ is a normalisation constant. 

In the case of a Graphical Gaussian model the joint probabilty $p$ can be written as
%
\begin{equation}
\label{eq:graphical_gaussian_models_2}
	p\left(x|\mathbf{\Theta}\right) = \frac{\left|\mathbf{\Theta}\right|^{1/2}}{\left(2\pi\right)^{N/2}}\exp\left(-\frac{1}{2} \mathbf{x}^T\mathbf{\Theta}\mathbf{x}\right),
\end{equation}
%
where $\mathbf{\Theta}$ is the inverse covariance or precision matrix
\end{comment}
%
In Gaussian Graphical models, the problem of finding the graph $\mathcal{G}$ is reduced to that of estimating a precision matrix $\mathbf{\Theta}$ that encodes the conditional relationship between the nodes. In the previous section, we saw that the production network influences the correlation of firms' growth $g_i$. If we consider each vector $\mathbf{g}\left(t\right)$ as a drawn from a joint probability distribution where the correlations are driven by the supply chain, Gaussian graphical models seem well equipped to reconstruct the production network if one ignores the fact that the growth rates do not have a Gaussian distribution.\footnote{Indeed, the marginal distribution of $x_i$ in Eq.\eqref{eq:graphical_gaussian_models_2} is clearly a Gaussian distribution.} We think nonetheless that, because the growth rates show a Gaussian-like central region, as shown by~\cite{growthrates}, it is reasonable to use this model to attempt a reconstruction.  

We therefore use the \textit{Graphical Lasso method} to construct an estimator $\widehat{\mathbf{\Theta}}$ of $\mathbf{\Theta}$ by solving the following optimisation problem:\footnote{
	This is the result of applying Bayes theorem assuming a constant prior for $\mathbf{\Theta}$.
}
\begin{equation}
\label{eq: graphical lasso}
	\widehat{\mathbf{\Theta}}=\text{argmax}_{\mathbf{\Theta}} \log\det\mathbf{\Theta} - \mathrm{tr}\left(\widehat{\mathbf{C}}\mathbf{\Theta}\right) - \alpha\|\mathbf{\Theta}\|_1,
\end{equation}
with $\widehat{\mathbf{C}} = \frac{1}{T}\mathbf{G}\mathbf{G}^T$ the sample covariance matrix, $\det\left(\cdot\right)$ the determinant and $\mathrm{tr}\left(\cdot\right)$ the trace. The first two terms can be thought of as the log-likelihood of $\mathbf{\Theta}$ in the Gaussian Graphical Model, while $\alpha \left|\mathbf{\Theta}\right|$ is an $L^1$ regularisation term with parameter $\alpha$. This approach will, in general, recover a matrix $\mathbf{\Theta}$ with both positive and negative entries. In this setting, a positive off-diagonal entry $\theta_{ij}$ of the precision matrix implies a negative partial correlation between  $\mathbf{x}_i$ and $\mathbf{x}_j$, whose interpretation is problematic since we would like $\mathbf{\Theta}$ to proxy the adjacency matrix of the network. 

References \cite{lake_tenenbaum_2010,daitch2009,chenhui2015} suggest instead searching for the precision matrix among the set $\mathcal{S}_{\mathbf{\Theta}}$ of possible Graph Laplacian matrices, 
\begin{equation}
\label{eq:laplacian_matrices}
	\mathcal{S}_{\mathbf{\Theta}} = \left\{\mathbf{\Theta} | \theta_{ij} = \theta_{ji} < 0\ \text{for}\ i\neq j, \theta_{ii}=-\sum_{j\neq i}\theta_{ij} \right\}.
\end{equation}

Conditioning $\widehat{\mathbf{\Theta}}$ to be in the set of possible graph Laplacians has two interesting consequences. First, the graph Laplacian $\mathbf{L}$ uniquely determines the adjacency matrix $\mathbf{W}$ of the graph; thus, the problem in \eqref{eq: graphical lasso} with the assumption $\mathbf{\Theta} \in \mathcal{S}_\mathbf{\Theta}$ creates a direct connection between the data and the topology of the network. Second, since the time series $\mathbf{g}_i$ has zero mean, we can write the trace $\left(\widehat{\mathbf{C}}\mathbf{\Theta}\right)$ as 
\begin{equation}
\label{eq:trace_to_smoothness}
    \mathrm{tr}\left(\widehat{\mathbf{C}}\mathbf{\Theta}\right) = \frac{1}{T}\mathrm{tr}\left( \mathbf{G}\mathbf{G}^T\mathbf{\Theta} \right) = \frac{1}{T}\sum_{i,j}\sum_{t=1}^T\theta_{ij}\left(g_i(t) - g_j(t)\right)^2.
\end{equation}
The term on the right hand of the equation measures the (squared) difference between the observation on firms $i$ and $j$ ($\mathbf{g}_i$ and $\mathbf{g}_j$), computed over couples of connected firms ($\theta_{ij} > 0$); it is generally known as the quadratic energy function and quantifies the \textit{smoothness} of $\mathbf{G}$ over the graph with Laplacian $\textbf{L}$. For an economic interpretation, the second term in \eqref{eq: graphical lasso}, $\mathrm{tr}\left(\widehat{\mathbf{C}}\mathbf{\Theta}\right)$, can be interpreted as a penalty term affecting networks over which $\mathbf{G}$ is not smooth, i.e., a production network that exhibits large differences between the growth rates of connected firms.

In \cite{Kumar_et_al_2019} (see Appendix \ref{app:network_reconstruction_algorithm}), the authors propose an efficient algorithm to solve the problem in Eq.\eqref{eq: graphical lasso} while also enforcing some (soft) constraints on the spectrum $\text{Sp}(\mathbf{\Theta})$ of the Laplacian matrix. The problem becomes
\begin{equation}
\label{eq:structured_gaussian_graphical_model}
	\begin{split}
    \widehat{\mathbf{\Theta}}=&\text{argmax}_{\mathbf{\Theta}} \log\det\mathbf{\Theta} - \mathrm{tr}\left(\widehat{\mathbf{C}}\mathbf{\Theta}\right) - \alpha\|\mathbf{\Theta}\|_1, \\
		&\textrm{subject to} \quad \mathbf{\Theta} \in \mathcal{S}_{\mathbf{\Theta}},\ \text{Sp}(\mathbf{\Theta})\subset \mathcal{S}_{\lambda} \\
	\end{split}
\end{equation}
where $\mathcal{S}_{\lambda}$ is the set of admissible spectra that we choose. Because the spectrum of the Laplacian encodes information about the underlying network's topology, choosing $\mathcal{S}_{\lambda}$ appropriately allows us to enforce high-level topological features on the reconstructed network.

We, therefore, attempt to use the algorithm provided in \cite{Kumar_et_al_2019} to reconstruct the production network. In the following, we assume that we know the network's density in advance and that we also have a reliable estimate for the number of links within and across different sectors. This information would not be available directly in a real-world situation, but the literature on production networks and other available data sources as input-output tables allow informed guesses (see, e.g., \cite{bacilieri2022}). This means that our results should be placed halfway between a proof of concept and a realistic use case.

\begin{figure}
\centering
    \includegraphics[width=.8\textwidth]{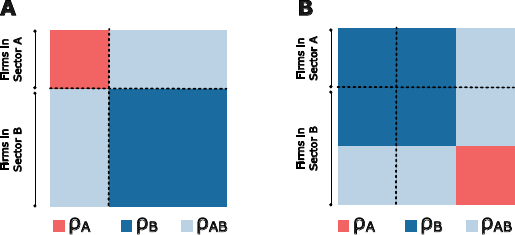}
    \caption{(A) A stylised representation of an adjacency matrix with two sectors. The density of links between the $n_A$ firms in sector A is $\rho_A$, the density of links between the $n_B$ firms in sector B is $\rho_B$, and the density of links across the two sectors is $\rho_{AB}$. (B) Another adjacency matrix. There are two group of firms of size $n_A$ (right bottom corner of the matrix) and $n_B$ (top left corner of the matrix). The density within firms in the first group is $\rho_B$, the density between firms in the second group is $\rho_A$, and the density across the groups is $\rho_{AB}$. The graph Laplacian of the matrix in (A) and that of the matrix in (B) will have the same spectrum. However, the density within and across sectors in (B) is different from that in (A).}
    \label{fig:matrix_explanation}
\end{figure}

We must however slightly modify this algorithm to apply it to our specific situation. Indeed, a problem with the algorithm described in \cite{Kumar_et_al_2019} is that, while it is possible to encode a given community structure by constraining the Laplacian, we are not able to specify which firms should go into which community (see Fig. \ref{fig:matrix_explanation}).

\begin{figure}
	\framebox{\includegraphics[width=\textwidth]{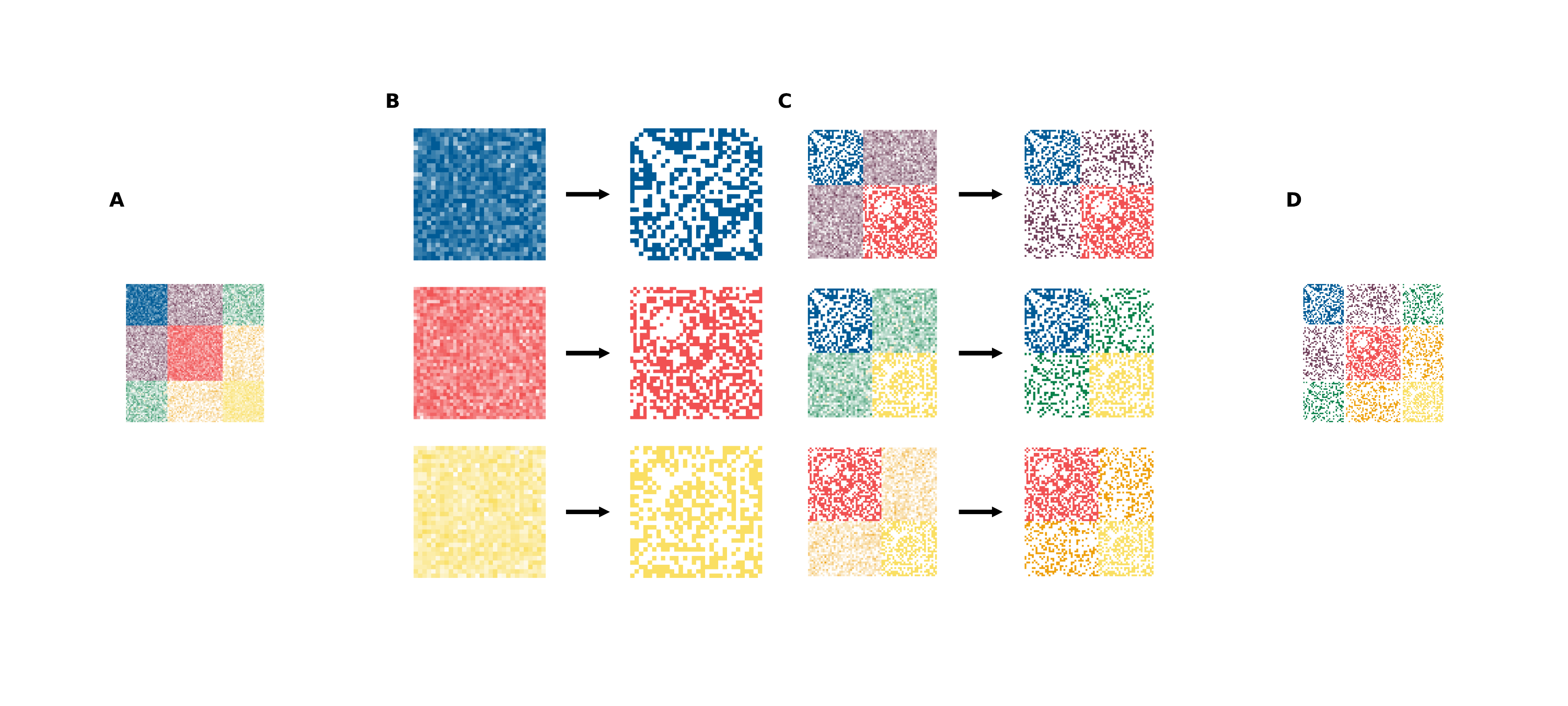}}
	\caption{Reconstruction of the supply chain networks. The original correlation matrix (A) is split into different industry sectors. First, we reconstruct the diagonal blocks (B). Then, we reconstruct the off-diagonal blocks (C). Finally, we re-assemble the blocks together (D).}
\label{fig:network_reconstruction}
\end{figure}

To solve this, we have devised the following procedure. First, we split $\widehat{\mathbf{C}}$ into diagonal and off-diagonal blocks based on firm industries. Next, we use the procedure defined in \eqref{eq:structured_gaussian_graphical_model} to reconstruct each diagonal block independently. Thirdly, we go through all the possible pairs of diagonal blocks and -- keeping the diagonal blocks equal to those that were reconstructed in the previous step -- we reconstruct the off-diagonal blocks. Finally, we assemble all the blocks together to obtain the entire adjacency matrix; this procedure is shown graphically in Fig. \ref{fig:network_reconstruction}.

Every time we reconstruct a network, we choose the parameter $\alpha$ to match the empirical network density. To reconstruct the diagonal blocks, we use the spectrum obtained by averaging over the spectra  1000 Erd\H{o}s-R\'enyi random networks' Laplacians, with probability $p$ equal to the desired density. Similarly, to reconstruct the off-diagonal blocks, we use the spectrum obtained by averaging over the spectra of 1000 block models' Laplacians, where the probabilities of links within and across each block are chosen to match the desired density.  We provide details on the reconstruction algorithm in Appendix \ref{app:network_reconstruction_algorithm}.

We ran our procedure over several different subparts of the real production network, each composed of a minimum of 300 to a maximum of 500 firms. We compared our results to those of two random benchmarks: an Erdos-Renyi graph and an industrial sector block model, built as in \ref{sec:network_correlation}. 
\begin{figure}
	\includegraphics[width=\textwidth]{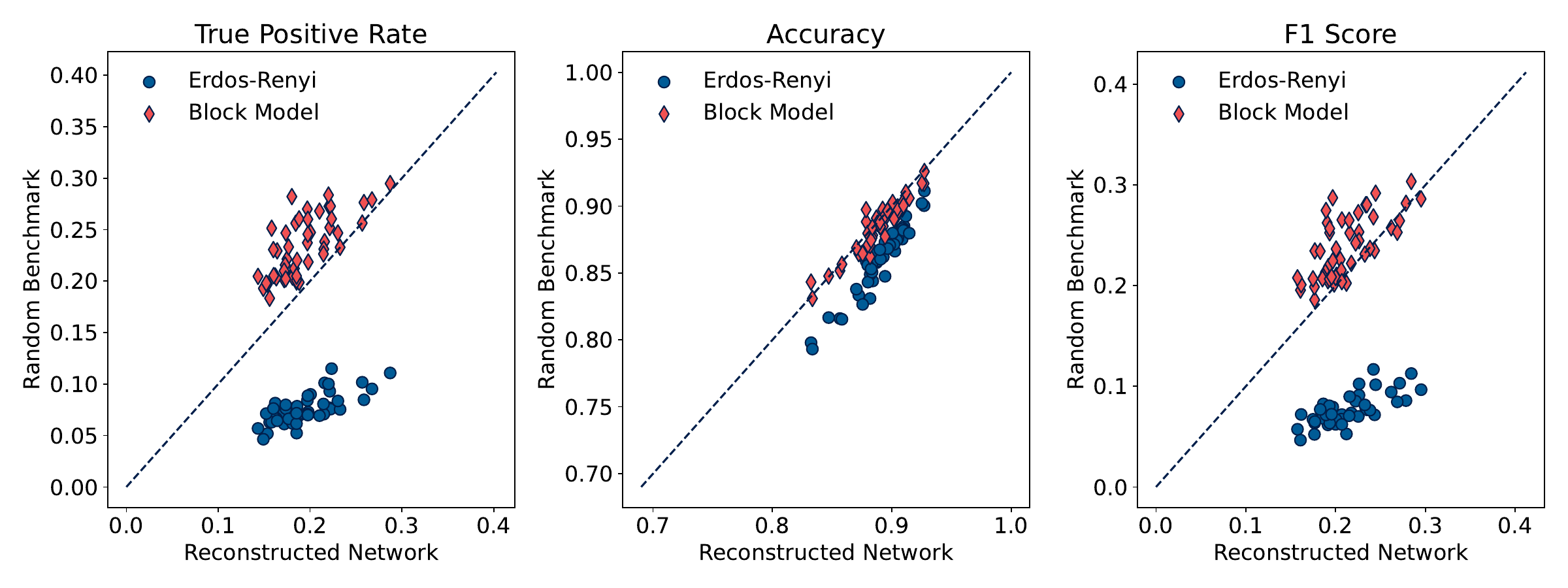}
	\caption{True Positive rate (left), Accuracy (middle), and F1 Score (right)of the reconstructed networks, plotted against the same metrics for the two different random benchmarks.}
	\label{fig:reconstruction_results}
\end{figure}
While our approach seems to have the highest accuracy, it fails to consistently beat the block model benchmark on the other metrics we tested (Fig. \ref{fig:reconstruction_results}).

\section{Conclusions}
\label{sec:conclusions}
In this paper, we studied if the correlation between firms' growth time series could be useful in reconstructing production networks. Using FactSet's supply chain network as a use case and several random network models as benchmarks, we have first shown that the growths of firms connected in the production networks are on average more correlated than those of randomly selected firms' pairs. We have shown that this effect fades gradually as one looks at the average correlation between pair of firms at an increasing network distance along the supply chain. Finally, we have framed the production network reconstruction in the context of graph learning and tested some recent techniques developed in the field to identify trade connections between firms. Our approach did not seem to significantly improve the benchmark, but we believe that it could still be improved to deliver good results. First, it relies on a mechanism that can be easily accepted as universal: the growth of business partners is correlated. Improvements in the estimation of these correlations, using techniques developed for financial data \citep{bun2017} and multiple time series (e.g., stock returns) will automatically improve our returns. Second, it is a fully "unsupervised" approach, which does not require the training of a model and is not prone to over-fitting. Third, it requires data that is easily accessible (firms' sales) and, to a certain extent, substitutable (e.g., we obtained similar results when we looked at the correlation of firms' stock returns). Finally, it generates a network that matches a set of desired topological features. This last point also highlights interesting avenues of research: as more "universal" production networks' features will be documented, and better generative models for these networks will be developed, the more effective our approach will be. 

% Main body ends here

\section*{Acknowledgements}
We would like to thank Jean-Philippe Bouchaud, Fran\c{c}ois Lafond, Doyne Farmer, and Xiaowen Dong for their numerous suggestions for this work, and Andrea Bacilieri for her help in handling the data. We would also like to thank the participants of the 2022 CSH–INET Workshop on Firm-Level Production Networks and the CCS 2022, in particular Christian Diem and Tobias Reisch, for the useful feedback, and Stefan Thurner and the network economics group at CSH Vienna for their hospitality and insight. This work was supported by Baillie Gifford and the Institute for New Economic Thinking at the Oxford Martin School.

\bibliography{biblio.bib}

\newpage
\appendix

% Appendix starts here

\section{Network Reconstruction Algorithm}
\label{app:network_reconstruction_algorithm}

The algorithm used to solve the problem in Eq.\eqref{eq:structured_gaussian_graphical_model} has first been proposed by ~\cite{Kumar_et_al_2019, Kumar_et_al_2020} in the context of structured Graph Learning. The authors formulate the problem as the following. Let $\boldsymbol{x} = \left[x_1, x_2, \ldots, x_p\right]^T$ be a $p$-dimensional, zero-mean, random vector (in the practical case, this would be the collection of the "cleaned" time series $\tilde{\mathbf{g}}_1, \ldots, \tilde{\mathbf{g}}_N$) associated with an undirected graph $\mathcal{G} = \left(\mathcal{V}, \mathcal{E}\right)$, where $\mathcal{V} = \left\{ 1, 2, \ldots, p \right\}$ is a set of nodes corresponding to the elements of $\boldsymbol{x}$, and $\mathcal{E} \in \mathcal{V} \times \mathcal{V}$ is the set of edges connecting nodes. In the \textit{Gaussian Graphical modeling} framework, learning a graph corresponds to solving the optimization problem
\begin{equation}
\label{eq:appendix_graph_learning_problem}
\max_{\Theta \in \mathcal{S}^p_{++}} \log \det\left(\Theta\right) - \text{tr}\left(\Theta S \right) - \alpha h\left(\Theta\right),
\end{equation}
where $\Theta \in \mathbb{R}^{p\times p}$ denotes the desired graph matrix, $\mathcal{S}^p_{++}$ denotes the set of $p \times p$ positive definite matrices, $S \in \mathbb{R}^{p \times p}$ is the covariance matrix obtained from the data, $S = \frac{1}{n}\boldsymbol{x}^T\boldsymbol{x}$, $h\left(\cdot\right)$ is a generic regularisation term, and $\alpha$ is a coefficient tuning the strength of the regularisation. As we saw in \ref{sec:reconstruction}, a matrix $\Theta \in \mathbb{R}^{p\times p}$ is called a combinatorial graph Laplacian matrix if it belongs to the set 
\begin{equation}
\label{eq:appendix_graph_laplacian_set}
\mathcal{S}_{\Theta} = \left\{\Theta | \theta_{ij} = \theta_{ji} < 0\ \text{for}\ i\neq j, \theta_{ii}=-\sum_{j\neq i}\theta_{ij} \right\}.
\end{equation}
The Laplacian Matrix $\Theta$ is a symmetric, positive semidefinite matrix with zero row sums. In the framework of network theory, a Laplacian matrix $\Theta$ is computed from a graph's adjacency matrix $A$ as $\Theta = D - A$, where $D$ is a diagonal matrix and $D_{ii}$ is the degree of node $i$. It is straightforward to see that the adjacency matrix of a graph can be recovered from the Laplacian matrix simply as $A = \Theta \odot \left(I -\mathbb{1}\right)$, where $I$ is the identity matrix, $\mathbb{1}_{ij} = 1$, and $\odot$ is the element-wise product. The structural properties of a graph are encoded in the eigenvalues of its Laplacian so that being able to constraint the spectrum of the matrix $\Theta$ in the optimization problem in Eq.\eqref{eq:appendix_graph_learning_problem} allows to enforce some structural constraints on the reconstructed network. The goal hence becomes that of solving the problem
\begin{equation}
\label{eq:appendix_graph_learning_problem_2}
\begin{aligned}
    \max_{\Theta} \quad & \log\text{gdet}\Theta - \mathrm{tr}\left(S\Theta\right) - \alpha h\left(\Theta\right), \\
        \textrm{subject to} \quad & \Theta \in \mathcal{S}_{\Theta},\ \mathbf{\lambda}\left(\Theta\right) \in \mathcal{S}_{\mathbf{\lambda}}, \\
    \end{aligned}
\end{equation}
where $\text{gdet}\left(\Theta\right)$ denotes the \textit{generalised determinant}\footnote{
    Note that in the main text, we have not made explicit the difference between $\text{gdet}\left(\Theta\right)$ and $\det \left(\Theta\right)$ to improve readability.
}
of the matrix $\Theta$, defined as the product of its non-zero eigenvalues, $\mathbf{\lambda}\left(\Theta\right)$ denotes the set of eigenvalues of $\Theta$, and $\mathcal{S}_{\lambda}$ is the set containing the spectral constraints on the eigenvalues. As the authors in \cite{Kumar_et_al_2019} point out, from the probabilistic perspective, if the data is generated from a multivariate Gaussian distribution $\mathcal{N}\left(0, \Theta^{\dagger}\right)$, then Eq.\eqref{eq:appendix_graph_learning_problem_2} can be viewed as a penalized maximum likelihood estimation of the structured precision matrix of an attractive Gaussian Markov Random Field model, while, if $\boldsymbol{x}$ is arbitrarily distributed, the problem in Eq.\eqref{eq:appendix_graph_learning_problem_2} corresponds to minimizing a penalized log-determinant Bregman divergence (a common measure of distance for probability distributions), and hence its solution should anyway result in a meaningful graph. In the main body of the paper, we saw how we assume to know the spectrum $\bar{\boldsymbol{\lambda}}$ of the target matrix is known, so we can define $\mathcal{S}_{\lambda}$ as
\begin{equation}
\label{eq:appendix_definition_S_lambda}
    \mathcal{S}_{\boldsymbol{\lambda}} = \left\{\lambda_i = \bar{\lambda}_i,\ \forall i \in \left[1,p\right]\right\}.
\end{equation}
To solve the optimisation problem in Eq.\eqref{eq:appendix_graph_learning_problem_2}, the authors in \cite{Kumar_et_al_2019} first introduce a \textit{Graph Laplacian linear operator} $\mathcal{L}$ to transform a generic, non-negative vector $\boldsymbol{w} \in \mathbb{R}_+^{p\left(p-1\right)/2}$ to a Laplacian matrix $\mathcal{L}\boldsymbol{w} \in \mathbb{R}^{p\times p}$. The linear operator $\mathcal{L}:\boldsymbol{w} \in \mathbb{R}_+^{p\left(p-1\right)/2} \rightarrow \mathcal{L}\boldsymbol{w} \in \mathbb{R}^{p\times p}$ is formally defined as
\begin{equation}
\label{eq:graph_laplacian_operator}
\left(\mathcal{L}\boldsymbol{w}\right)_{ij} = 
    \begin{cases}
        -w_{i+d_j} & i> j, \\
        \left(\mathcal{L}\boldsymbol{w}\right)_{ji} & i < j,  \\
        \sum_{i \neq j} \left(\mathcal{L}\boldsymbol{w}\right)_{ij} & i=j, \\
    \end{cases}
\end{equation}
where $d_j = -j + \frac{j-1}{2}\left(2p-1\right)$. The adjoint operator $\mathcal{L}^*: Y \in \mathbb{R}^{\left(p \times p\right)} \rightarrow \mathcal{L}^*T \in \mathbb{R}^{\frac{p\left(p-1\right)}{2}}$ is derived to satisfy $\left<\mathcal{L}\boldsymbol{w}, Y\right> = \left<\boldsymbol{w}, \mathcal{L}^*Y\right>$. While the definition of the two operators might seem cumbersome at first glance, their interpretation is fairly straightforward
%: given a Laplacian matrix $Y$, the operator $\mathcal{L}^*$ flattens the upper-triangular part of $-Y$ into a vector $\boldsymbol{w}$; $\mathcal{L}$ inverts the process 
(see Fig, \ref{fig:laplacian_operator}).
\begin{figure}[tb]
\centering
    \begin{tikzpicture}
    % Left part
    \matrix (A) [matrix of nodes, left delimiter=(, right delimiter=)]{
        $w_1$, & $w_2$, & $w_3$, & $w_4$, & $w_5$, & $w_6$ \\
    };
    % Central part
    \draw [-{Latex[length=5mm, width=5pt]}] (2.8, 0.8) -- node[above=1mm] {$\mathcal{L}$} (4.0,0.8);
    \draw [-{Latex[length=5mm, width=5pt]}] (4.0, -0.8) -- node[below=1mm] {$\mathcal{L}^{-1}$} (2.8, -0.8);
    % Right part
    \matrix (B) at (8.5, 0) [matrix of nodes,  left delimiter=(, right delimiter=)]{
        $\sum_{i \in \left\{1, 2, 3\right\}}w_i$ & $-w_1$ & $-w_2$ & $-w_3$\\
        $\ldots$ & $\sum_{i \in \left\{1, 4, 5\right\}}w_i$ & $-w_4$ & $-w_5$\\
        $\ldots$ & $\ldots$ & $\sum_{i \in \left\{2, 4, 6\right\}}w_i$ & $-w_6$\\
        $\ldots$ & $\ldots$ & $\ldots$ & $\sum_{i \in \left\{3, 5, 6\right\}}w_i$\\
    };
    \end{tikzpicture}
    \caption{Given a Laplacian matrix $Y$, the operator $\mathcal{L}^{-1}$ flattens the upper-triangular part of $-Y$ into a vector $\boldsymbol{w}$. $\mathcal{L}$ inverts the process.}
\label{fig:laplacian_operator}
\end{figure}
\begin{figure}[tb]
\centering
    \begin{tikzpicture}
    % Left part
    \matrix (A) [matrix of nodes, left delimiter=(,right delimiter=)]{
        $y_{11}$ & $y_{12}$ & $y_{13}$ & $y_{14}$ \\
        $y_{21}$ & $y_{22}$ & $y_{23}$ & $y_{24}$ \\
        $y_{31}$ & $y_{32}$ & $y_{33}$ & $y_{34}$ \\
        $y_{41}$ & $y_{42}$ & $y_{43}$ & $y_{44}$ \\
    };
    % Central part
    \draw [-{Latex[length=5mm, width=5pt]}] (2.8, 0.0) -- node[above=1mm] {$\mathcal{L}^*$} (4.0,0.0);
    % Right part
    \matrix (B) at (7.0, 0) [matrix of nodes,  left delimiter=(,right delimiter=)]{
        $y_{11} - y_{21} - y_{12} + y_{22}$ \\
        $y_{11} - y_{31} - y_{13} + y_{33}$ \\
        $y_{11} - y_{41} - y_{14} + y_{44}$ \\
        $y_{22} - y_{32} - y_{23} + y_{33}$ \\
        $y_{22} - y_{42} - y_{24} + y_{44}$ \\
        $y_{33} - y_{43} - y_{34} + y_{44}$ \\
    };
    \end{tikzpicture}
    \caption{The adjoint operator $\mathcal{L}^*$ transforms a symmetric matrix in a vector. Above, an example for a $4 \times 4$ matrix.}
\label{fig:laplacian_transpose_operator}
\end{figure}
The Laplacian operator $\mathcal{L}$ allows reformulating the optimization problem in a simpler way. First, by the definition of $\mathcal{L}$, the set of constraints in Eq.\eqref{eq:appendix_graph_laplacian_set} can be expressed as $\mathcal{S}_{\Theta} = \left\{ \Theta = \mathcal{L}\mathbf{w} | \mathbf{w} \geq 0 \right\}$. Second, if we choose $h\left(\Theta\right)$ to be the $\mathcal{L}_1$-regularisation function, since $\left(\mathcal{L}\boldsymbol{w}\right)_{ij} < 0$ for $i \neq j$ and $\left(\mathcal{L}\boldsymbol{w}\right)_{ij} > 0$ for $i = j$, the regularisation term $\alpha h\left(\mathcal{L}\boldsymbol{w}\right) = \alpha \|\mathcal{L}\boldsymbol{w}\|_1$ can be written as $\text{tr}\left(\mathcal{L}\boldsymbol{w}H\right)$, where $H = \alpha \left(2I - \mathbb{1}\right)$, which implies 
\begin{equation}
\label{eq:app_trace_transformation}
    \text{tr}\left(\mathcal{L}\boldsymbol{w}S\right) + \alpha h \left(\mathcal{L}\boldsymbol{w}\right) = \text{tr}\left(\mathcal{L}\boldsymbol{w}K\right),
\end{equation}
where $K = S + H$. We can now reformulate Eq.\eqref{eq:appendix_graph_learning_problem_2} as
\begin{equation}
\label{eq:appendix_graph_learning_problem_3}
    \begin{aligned}
    \min_{\boldsymbol{w}, U} \quad & - \log\text{gdet}\left(U\text{Diag}\left(\bar{\boldsymbol{\lambda}}\right)U^T\right) + \text{tr}\left(\mathcal{L}\boldsymbol{w}K\right) + \frac{\beta}{2}\|\mathcal{L}\boldsymbol{w} - U\text{Diag}\left(\bar{\boldsymbol{\lambda}}\right)U^T\|_F^2, \\
        \textrm{subject to} \quad & \boldsymbol{w} > 0, U^TU= I. \\
    \end{aligned}
\end{equation}
where $\mathcal{L}\boldsymbol{w}$ is the Laplacian matrix that we would like to decompose as $\mathcal{L}\boldsymbol{w} = U\text{Diag}\left(\bar{\boldsymbol{\lambda}}\right)U^T$, $\text{Diag}\left(\bar{\boldsymbol{\lambda}}\right) \in \mathbb{R}^{p\times p}$ is a diagonal matrix containing $\left\{\bar{\lambda}_i\right\}$ on its diagonal, and $U \in \mathbb{R}^{p\times p}$ is an orthogonal matrix. The constraints on the spectrum of the reconstructed matrix are enforced (softly) thanks to the spectral penalty term $\frac{\beta}{2}\|\mathcal{L}\boldsymbol{w} - U\text{Diag}\left(\bar{\boldsymbol{\lambda}}\right)U^T\|_F^2$. It is well known that every Laplacian matrix $\Theta$ will have at least one eigenvalue equal to zero, since $\Theta\cdot\mathbb{1} = 0$ by definition. Consequently, when solving \eqref{eq:appendix_graph_learning_problem_3}, the first eigenvalue and the corresponding eigenvector can be dropped from the optimization formulation. Now $\bar{\boldsymbol{\lambda}}$ only contains $q = p-1$ non zero eigenvalues in increasing order, $\left\{\lambda_j\right\}_{j=2}^p$; we can replace the generalized determinant in \eqref{eq:appendix_graph_learning_problem_3} with the standard determinant on $\text{Diag}\left(\bar{\boldsymbol{\lambda}}\right)$, and redefine $U$ as $U \in \mathcal{R}^{p \times q}$, containing the eigenvectors corresponding to non-zero eigenvalues in the same order. The orthogonality constraint becomes $U^T U = I_q$. In ~\cite{Kumar_et_al_2019}, the authors show how the problem in \eqref{eq:appendix_graph_learning_problem_3} can be solved with an iterative approach. If we define the vector $\boldsymbol{c}$,
\begin{equation}
\label{fig:appendix_vector_c}
\boldsymbol{c} = \left[\mathcal{L}^*\left(U\text{Diag}\left(\bar{\boldsymbol{\lambda}}\right)U^T - \frac{1}{\beta}K\right)\right],
\end{equation}
and the function $f\left(\boldsymbol{w}\right)$,
\begin{equation}
\label{fig:appendix_function_f}
f\left(\boldsymbol{w}\right) = \frac{1}{2}\|\mathcal{L}\boldsymbol{w}\|^2_F - \boldsymbol{c}^T\boldsymbol{w},
\end{equation}

at each step $t$, we can update $\boldsymbol{w}$ and $U$, as
\begin{equation}
\label{eq:appendix_updating_rule_w}
        \boldsymbol{w}^{t+1} = \left[\boldsymbol{w}^t - \frac{1}{2p}\nabla f\left(\boldsymbol{w}^t\right)\right]^+,
\end{equation}
\begin{equation}
    U^{t+1} = \Lambda\left(\mathcal{L}\boldsymbol{w}\right)[2: p],
\end{equation}
where $\Lambda\left(\mathcal{L}\boldsymbol{w}\right)$ is the matrix of the eigenvectors of $\mathcal{L}\boldsymbol{w}$, sorted by the corresponding eigenvalue. The algorithm can be run until convergence, $\boldsymbol{w}^{t+1} = \boldsymbol{w}^{t} = \boldsymbol{w}^*$, and the vector $\boldsymbol{w}^*$ can be used to reconstruct the Laplacian $\Theta = \mathcal{L}^*\boldsymbol{w}^*$, and the corresponding adjacency matrix. To reconstruct off-diagonal blocks of our Laplacian matrix, we have, at each iteration step, only updated the components of $\boldsymbol{w}$ corresponding to off-diagonal blocks, and again run the algorithm until convergence. While there is no theoretical guarantee that the algorithm will converge to the optimal solution of the optimization problem, our results suggest that this approach is still effective in reconstructing the network.

\section{Dataset construction}
\label{app:dataset_construction}

For the purposes of this paper, we accessed three different FactSet products: \textit{Standard Datafeed - Fundamentals V3 - Advanced - Global}, \textit{Standard Datafeed - Supply Chain relationship}, and \textit{APB - Standard Datafeed - Supply Chain Shipping Transaction}. We parsed information on companies' fundamentals (sales, market capitalization, capital expenditures, industrial sector, and geography) from the first dataset and used the other two to identify supply chain relationships.

\paragraph{Fundamentals} The fundamentals dataset is built from the following FactSet files:

\begin{enumerate}
\item Fundamentals
    \begin{itemize}
        \item \texttt{ff\_basic\_eu\_v3\_full\_5315/ff\_basic\_af\_eu.txt}
        \item \texttt{ff\_advanced\_eu\_v3\_full\_4524/ff\_advanced\_af\_eu.txt}

        \item \texttt{ff\_basic\_ap\_v3\_full\_5276/ff\_basic\_af\_ap.txt}
        \item \texttt{ff\_advanced\_der\_ap\_v3\_full\_4460/ff\_advanced\_der\_af\_ap.txt}

        \item \texttt{ff\_basic\_am\_v3\_full\_5258/ff\_basic\_af\_am.txt}
        \item \texttt{ff\_advanced\_der\_am\_v3\_full\_4484/ff\_advanced\_der\_af\_am.txt}
    \end{itemize}

\item FX Rates
    \begin{itemize}
        \item \texttt{fx\_rates\_usd.txt}
    \end{itemize}

\item Symbology
    \begin{itemize}
        \item \texttt{sym\_hub\_v1\_full\_9915/sym\_coverage.txt}
        \item \texttt{sym\_hub\_v1\_full\_9915/sym\_entity\_sector.txt}
        \item \texttt{f\_sec\_hub\_v3\_full\_5299/ff\_sec\_entity\_hist.txt}
    \end{itemize}

\end{enumerate}

The \textit{Fundamentals} files contain the (yearly) information regarding companies' sales, number of employees, and r\&d expenses, and a \textit{currency} column that states the features' currency. We can convert all of these features into USD using the FX Rates table provided by FactSet. The original fundamentals files are not at the \textit{security} level, not at the company's one. To create a dataset at the company level, FactSet provided us with the following example query,

\begin{quote}
\texttt{
    Select a.factset\_entity\_id, c.fsym\_id,c.date,c.ff\_sales \\
    from [sym\_v1].[sym\_sec\_entity] a\\
    join [sym\_v1].[sym\_coverage] b on a.fsym\_id = b.fsym\_id \\
    join [ff\_v3].[ff\_basic\_qf] c on c.fsym\_id = b.fsym\_regional\_id \\
    where a.factset\_entity\_id ='05HK0W-E'and a.fsym\_id = b.fsym\_primary\_equity\_id \\
    },
\end{quote}
that we "translated" to Python. We used \texttt{sym\_hub\_v1\_full\_9915/sym\_entity\_sector.txt} to assign the correct SIC code to each of the firms.

\paragraph{Supply Chain edgelist}

The Supply Chain's edge list is built from the following FactSet files:
\begin{enumerate}
    \item Supply Chain
    \begin{itemize}
        \item \texttt{ent\_supply\_chain\_v1\_full\_2354/ent\_scr\_supply\_chain.txt}
    \end{itemize}
    \item Shipments
    \begin{itemize}
        \item \texttt{sc\_ship\_trans\_current\_v1\_full\_1146/sc\_ship\_trans\_curr\_1.txt}
        \item \texttt{sc\_ship\_trans\_current\_v1\_full\_1146/sc\_ship\_trans\_curr\_2.txt}
        \item \texttt{sc\_ship\_trans\_current\_v1\_full\_1146/sc\_ship\_trans\_curr\_3.txt}
        \item \texttt{sc\_ship\_trans\_current\_v1\_full\_1146/sc\_ship\_trans\_curr\_4.txt}
    \end{itemize}
    \item Mappings
    \begin{itemize}
        \item \texttt{ent\_entity\_advanced\_v1\_full\_6896/factset\_entity\_structure.csv}
        \item \texttt{sc\_ship\_trans\_hub\_v1\_full\_1120/sc\_ship\_parent.txt}
    \end{itemize}
\end{enumerate}

The Supply Chain and Shipment files both contain an edge list (supplier-to-customer and shipper-to-consignee respectively). The mapping files have two columns “FACTSET\_ENTITY\_ID” and “FACTSET\_ULT\_PARENT\_ENTITY\_ID”. We assume that every FACTSET\_ENTITY\_ID that is not present in the mapping is an ultimate parent company.

\paragraph{Coordinates} The firms' geographical position was fetched from the following files:
\begin{enumerate}
    \item FactSet's Addresses
    \begin{itemize}
        \item \texttt{ent\_supply\_chain\_hub\_v1\_full\_2355/ent\_scr\_address.txt}
        \item \texttt{sc\_ship\_trans\_hub\_v1\_full\_1120/sc\_ship\_address\_coord.txt}
        \item \texttt{sym\_hub\_v1\_full\_9915/sym\_address.txt'}
    \end{itemize}
\end{enumerate}

\section{Other cleaning strategies}
\label{app:cleaning}

While working on the paper, we tested two other methods to process the correlation matrix in a way to maximize the gap between the average correlation along the supply chain and those of the random benchmarks (see \ref{sec:network_correlation}). After cleaning the market mode, we tried to see whether we could remove some sector-specific trends from the time series. For each industrial sector $\alpha$ we defined the quantity 
$$
s_{\alpha}\left(t\right) = \sum_{i, i \in \alpha} x_{i}\left(t\right),
$$
where $x_i\left(t\right)$ is the growth time series of firm $i$, and the sum runs on all the firms in sector $alpha$. We assumed that we could write the time series $x_i\left(t\right)$as
$$
x_i\left(t\right) = \xi_i\left(t\right) + k_i s_{\alpha}\left(t\right),
$$
We estimated the coefficient $k_i$ as the correlation between $x_i$ and $s_{\alpha}$, and cleaned the time series by computing the difference
$$
\xi_i(t) = x_i\left(t\right) - \hat{k}_i s_{\alpha}(t),
$$
where $\hat{k}_i$ is the estimated value for $k_i$. 

We also investigated if more signals could be extracted by considering lags between firms' time series. We defined the lagged correlation matrix $C\left(\tau\right)$, defined as 
$$
C\left(\tau\right) = \avg_t\left[x_i(t)x_j(t+\tau)\right],
$$
and its symmetrised version $C'\left(\tau\right)$ as
$$
C'(1) = \frac{1}{2}\left[C(1) + C(-1)\right].
$$
We then computed a linear combination $\left[C'(0) + C'(1)\right]$, and computed the average value of this matrix over the supply chain and the random benchmarks.

None of the two approaches improved significantly the outcomes we discussed. However, we can't exclude that a more thorough investigation of these techniques, their combination, and the analysis of other time series (e.g., firms' market returns) could improve the results of this paper.

% Appendix ends here

\end{document}